\documentclass[preprint,showpacs,pre,a4paper]{revtex4}

\usepackage[latin1]{inputenc}
\usepackage{graphicx}

\newcommand{\gcpf}{\ensuremath{\mathcal{Z}}}

\begin{document}

\title{The Bethe approximation for the hydrogen-bonding self-avoiding walk in a solvent}

\author{D P  Foster}
\author{M Aniambossou}

\affiliation{Laboratoire de Physique Th\'eorique et Mod\'elisation
(CNRS UMR 8089), Universit\'e de Cergy-Pontoise, 2 ave A. Chauvin
95302 Cergy-Pontoise cedex, France}

\begin{abstract}
A square-lattice model for the formation of secondary structures in proteins, the hydrogen-bonding model,  extended to include the effects of solvent quality, is examined in the framework of the Bethe approximation.
\end{abstract}


\maketitle

\section{Introduction}
There is a long tradition of using lattice models of polymers in an attempt to capture the essential features of the physics of polymers in solution\cite{Gennes-P-G:1979sh,Cloiseaux:1990mi,Vanderzande:1998ce}.  The canonical lattice model is the self-avoiding walk (SAW) model extended by including interactions between nearest-neighbour visited sites on the lattice. These interactions model the quality of the solvent in which the polymer lies\cite{Domb:1974cq,Wall:1961tf}. When only one walk is considered, the model is thought to describe the behaviour of polymers in dilute solution. At high temperature the polymer
is `happy' in solution, whilst at low temperature the polymer collapses and precipitates from solution. These two regimes are separated by the $\Theta$-transition\cite{Flory:1971px,Gennes:1972xh,Gennes:1975}. In what follows, we refer to this model as the $\Theta$ model.

In the early 1990s  a model was introduced to model the effects of hydrogen bonding on the formation of secondary structures resulting from the folding of a protein: the Hydrogen-Bonding self-avoiding walk, introduced by Bascle, Garel and  Orland\cite{Bascle:1993fr}.  
In this model the presence of hydrogen bonds, essential in real proteins, was modelled by the 
presence of interactions between parallel 
straight sections of the walk, as shown in figure~\ref{hint}. They studied the model in the Hamiltonian Walk limit, where all the sites of the lattice are visited exactly once. The model was extended first to all densities\cite{Foster:2001xd}
and later to allow for solvent effects 
by including all the interactions present in the $\Theta$-point model, but with different interaction strengths, depending on the configuration of the walk
(again see figure~\ref{hint})\cite{Foster:2008fk}.

Exploring the phase diagrams of such  frustrated self-avoiding walk models is not an easy task, in most cases requiring good quality numerical methods. This is mainly due to the frustration effects intrinsic to the presence of interactions between portions of the walk which may be arbitrarily far apart along the walk. The standard Monte-Carlo methods applied to these models consist in studying 
finite walks on the infinite lattice and examining the finite-size behaviour of the walk. The current favourite Monte-Carlo methods are the PERM method\cite{PERM1,PERM2,PERM3,PERM4,PERM5}, the flat-PERM method\cite{ FP} and the parallel tempering method\cite{PT}. 
This limits the
use of the method to phase transitions coming from the zero-density high-temperature phase (the self-avoiding walk phase) but does not permit the study of phase transitions between different dense phases.
The use of Monte Carlo in such phases is extremely difficult, normally requiring the relaxation of 
some constraints, as is the case for the fluctuating-bond method\cite{binder94,Landau:2005fk}. This however tends to erase the very effects we wish to study.

Another approach which has proved useful in studying such models is the use of transfer matrices in which the partition and correlation functions are expressed in terms of products of matrices enabling numerically exact calculations on infinitely long strips of finite width\cite{Foster:2001xd,ds85}. This method
has two limitations: in practical terms the method is restricted to two dimensions, and 
the number of available widths is limited. The latter restriction may be alleviated by using the CTMRG method (corner transfer matrix re-normalisation group method), which enables calculations for much
larger lattice sizes\cite{Foster:2003mb,Foster:2003sh}. These methods enable the investigation of the phase-diagram in the entire phase-space.  

The hydrogen-bonding self-avoiding walk in a solvent was 
recently studied in two dimensions using a combination of transfer matrix and CTMRG methods\cite{Foster:2008fk}. It has also been independently studied using a modified PERM Monte-Carlo method in two and three 
dimensions\cite{Krawczyk:207qy}. 

All numerical methods are open to possible misinterpretation or artefacts, particularly when applied to models which include 
significant frustration effects. It is important to have some independent confirmation that the results obtained are reasonable. In this article we propose to provide such an independent
confirmation by performing a mean-field type calculation in
the form of the Bethe approximation for the hydrogen-bonding self-avoiding walk 
in a solvent and compare the results with the previously obtained numerical results. 

In the next section we present the model in detail. In section~\ref{bethe} we apply the Bethe approximation to our model and in section~\ref{results} we present our results. We finish with discussion
and conclusions in section~\ref{discussion}.

\section{Model}
\begin{figure}
 \includegraphics[width=12cm]{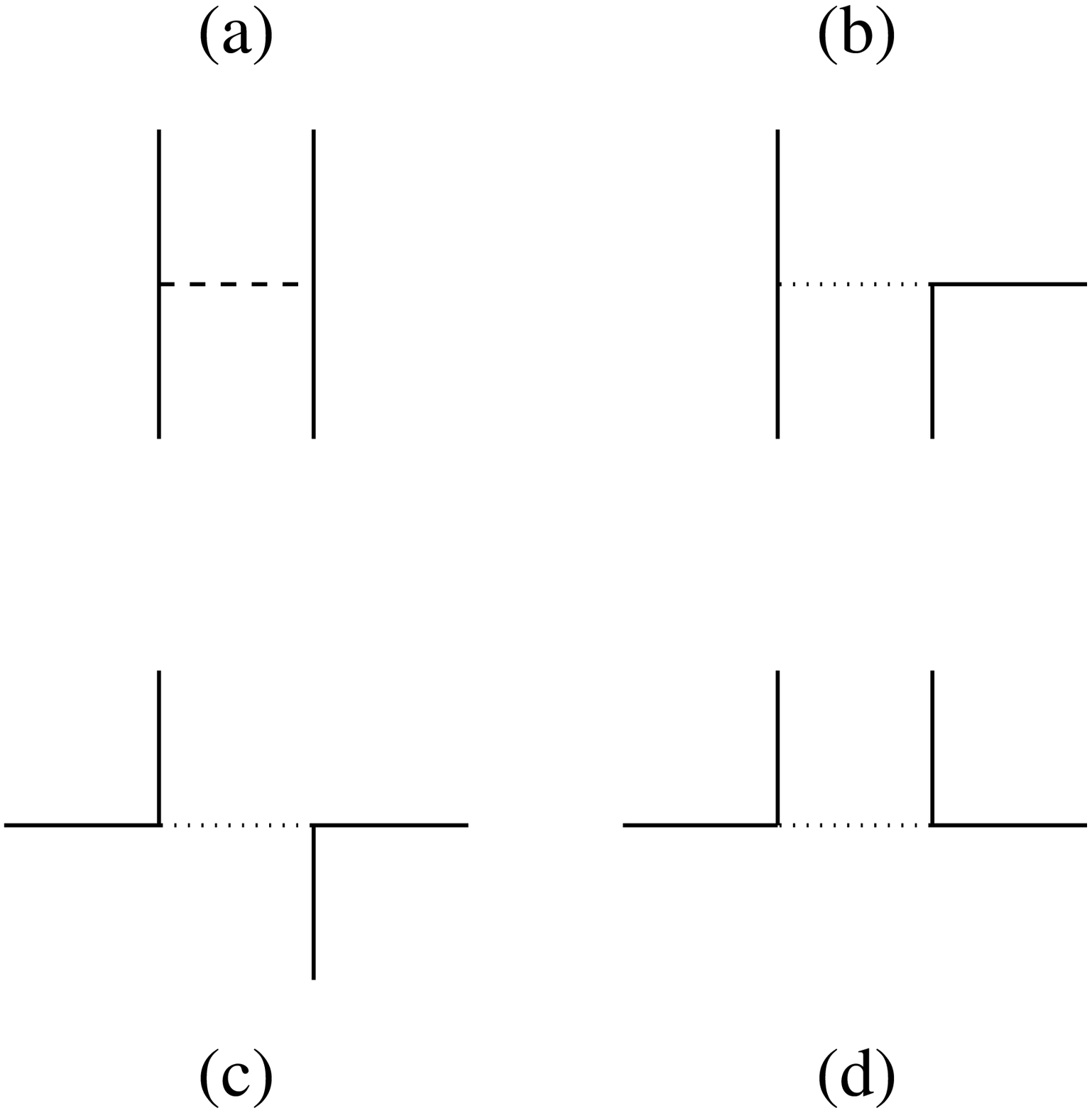}
 \caption{The nearest-neighbour interactions are split into two classes, those of type (a) where four bonds forming two parallel lines model the hydrogen bonds, whilst the others (b, c and d) model the solvent-mediated
 interactions. Configuration (a) induces a preferred orientation, whilst the other configurations do not.}\label{hint}
 \end{figure}

The model studied in this article involves the self-avoiding walk on the square lattice with interactions between non-consecutive visited nearest-neighbour sites on the lattice. 
In the standard $\Theta$-point model the nearest neighbour interactions model effective
interactions mediated by the solvent, given by the difference between
the monomer-monomer and monomer-solvent affinities. These interactions are isotropic. In the current model, the interactions are split into two sets, as shown in figure~\ref{hint}; those which specify a particular direction, the hydrogen bonds, and those that do not, which we shall refer to as the solvent-mediated interactions. Hydrogen bonds carry an interaction energy $-\varepsilon_H$ 
and the others carry an interaction energy $-\varepsilon$. The thermodynamic behaviour may be investigated by introducing the grand-canonical partition function, \gcpf, from which many of the 
relevant thermodynamic quantities may be calculated. The grand-canonical partition function is given by:
\begin{equation}
\gcpf=\sum_{\rm walks} K^N\exp\left(\beta\left(N_I\varepsilon+N_H\varepsilon_H\right)\right),
\end{equation}
where $N_I$ are the number of solvent-mediated interactions, and $N_H$ are the number of hydrogen bonds. The fugacity, which controls the average length of the walk, is denoted by $K$, and $N$ is the total length of the walk. For convenience we define $\alpha=\varepsilon/\varepsilon_H$, and 
without changing the physics of the model, we may set $\varepsilon_H=1$; this simply sets the temperature scale.  The partition function then becomes:
\begin{equation}
\gcpf=\sum_{\rm walks} K^N\exp\left(\beta\left(N_H+N_I\alpha\right)\right),
\end{equation}
from which we may calculate, for example, the free energy per site, $f$: 
\begin{equation}
\beta f=-\frac{1}{\Omega}\ln \gcpf
\end{equation}
where $\Omega$ is the number of lattice sites. The free energy as defined here is also the grand potential for the model in which we concentrate on the walk (rather than the lattice) and view the 
problem as grand canonical since the number of steps in the walk varies. Clearly the basic unit of the calculation is the lattice site, and so we choose the convention of referring to $f$ as the free energy per site, which corresponds to the standard picture in Bethe approximation calculations.

 The fugacity $K$ controls the average length of the walk. The average number of steps is given by
 \begin{equation}
 \langle N\rangle=K\frac{\partial \ln\gcpf}{\partial K}.
 \end{equation}
 The average length increases as $K$ is increased. 
 For fixed $\alpha$, if $\beta$ is small enough, then the average length diverges continuously 
 as $K$ approaches some critical value $K_c(\alpha,\beta)$. This defines the self-avoiding walk line, which extends to a (half) plane as $\alpha$ is varied.
If, on the other hand, $\beta$ is large enough then the average length jumps discontinuously at some 
value of the fugacity, $K=K^*(\alpha,\beta)$. 
Together these two regimes define the plane $K_\infty(\alpha,\beta)$ on which the walk length first diverges. This plane separates the high-temperature, zero-density,
 phase from the low-temperature, dense, phases.
 
\section{The Bethe approximation}\label{bethe}

\begin{figure}
\begin{center}
\includegraphics[width=12cm]{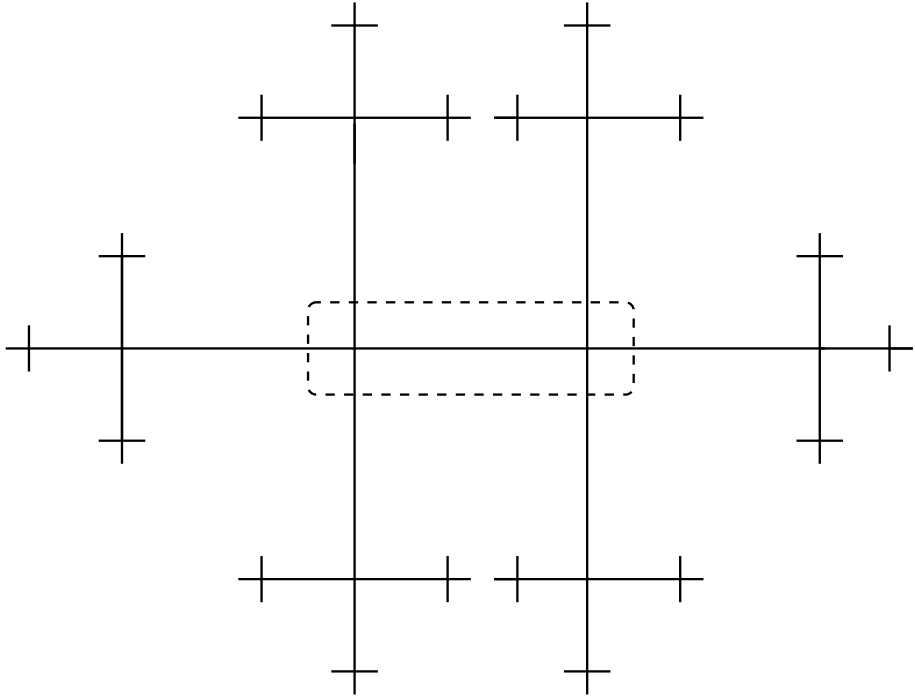}
\end{center}
\caption{The Bethe lattice representation of the  two-dimensional lattice. The dotted box shows the central bond, exhibiting the desired square-lattice geometry. }\label{blat}
\label{Bethe}
\end{figure}

In this section we describe briefly the Bethe approximation. For a good discussion of the Bethe approximaton see Ref. \cite{Baxter:1982gf}.
 The model of interest 
 is studied on the infinite Bethe lattice chosen to have the correct local geometry. The lattice 
 chosen for the square lattice is shown in figure~\ref{blat}. 
 The Bethe lattice is a hierarchical lattice built recursively from a central bond by adding to each 
 extremity $k$ new bonds. To each
dangling bond we add $k$ more bonds, and so on, such that no loops are formed.  Due to the hierarchical nature of the lattice, it is possible to build up expressions for the partition function recursively. To see this, it is convenient to consider the lattice as being divided into two branches,
left and right for the example shown in figure~\ref{blat}. We may introduce the partial partition functions
$W^{\rm l}_\sigma$ and $W^{\rm r}_\sigma$ for the left and right-hand branch, respectively. These 
partition functions are conditional upon the state $\sigma$ of the central bond. 
In our model there are four possible states: 
\begin{enumerate}
\item empty (state 0),
\item occupied with a link of the walk (state $K$),
\item occupied with a $\Theta$ interaction (state $\Theta$), and
\item occupied with a hydrogen bond (state $H$).
\end{enumerate}
By symmetry, the left and right branches will have the same partial partition functions, and so the l,r designation will be dropped. Each branch may be sub-divided into $k$ sub-branches, such that the $W_\sigma$ may be expressed in terms of the partial partition functions of the sub-branches. 
This procedure may be continued until the boundary bonds are reached. 
In order to do this explicitly, it is convenient to introduce the notion of the `generation' of a link, $n$,
 which is simply the distance of the link from the boundary. As a concrete example, consider the calculation of $W_K^{(n)}$, the partial partition function conditional on the central bond being occupied
 by a link of the walk.
 We must consider all the configurations on the bonds of the generation $(n-1)$, of which there are three for the 2 dimensional square lattice example shown in figure~\ref{blat}, which are compatible with the occupied central bond. Clearly there must be a bond leaving in one of the three directions, the other two bonds may be empty or occupied by a solvent-mediated interaction. If the bonds on the central bond and at generation $(n-1)$ line up, then the``empty" bonds may be occupied by hydrogen interactions. The weight $W_K^{(n)}$ is simply the sum of the Boltzmann weights corresponding to all these configurations, multiplied by the weight for adding the central link. To avoid the divergence of the
 partial partition functions it is 
 convenient to introduce normalised partition functions 
 $w^{(n)}_\sigma=W^{(n)}_\sigma/q_n$\cite{Pretti:2006qb} with $q_n$ chosen such that:
\begin{equation}
\sum_\sigma w^{(n)}_\sigma=1.
\end{equation}

This leads to recursion relations for the (normalised) partial partition functions:
\begin{equation}
w^{(n)}_\sigma=\frac{\lambda_\sigma}{q_n}\sum_{\{\gamma_i\}} C_{\sigma,\{\gamma_i\}}\prod_{i=1}^k w^{(n-1)}_{\gamma_i},
\end{equation}
where $\{\gamma_i\}$ is the set of state of the $k$ links forming generation $n-1$,  $\lambda_\sigma$ is the Boltzmann weight of the bond added at generation $n$, and the factor $C_{\sigma,\{\gamma_i\}}=1$ 
if the choice of the states $\{\gamma_i\}$ is compatible with the central state $\sigma$, and zero otherwise. 

It is known that there is no phase transition on the infinite Bethe lattice, since the number 
of boundary sites 
grows too rapidly. 
However the recursion relations may be used in the centre of the lattice as self-consistency equations for the two point mean-field theory for the corresponding square lattice. In this case, we assume we have translational invariance, and drop the generational superscripts. The equilibrium states are then given by solutions of the following set of recursion relations:
\begin{eqnarray}\label{rec1}
w_0&=&\frac{1}{q} \left\{w_0^3+\left(3(w_0+w_\Theta)+w_H\right)w_K^2\right\}\\\label{rec2}
w_K&=&\frac{K}{q} w_K\left(3(w_0+w_\Theta)^2+2(w_0+w_\Theta)w_H+w_H^2\right)\\\label{rec3}
w_\Theta&=&\frac{(e^{\alpha\beta}-1)}{q} w_K^2\left(3(w_0+w_\Theta)+w_H\right)\\\label{rec4}
w_H&=&\frac{(e^\beta-e^{\alpha\beta})}{q} w_K^2\left(w_0+w_\Theta+w_H\right)\\\nonumber\label{rec5}
q&=&w_0^3+K\left(3(w_0+w_\Theta)^2
+(2(w_0+w_\theta)+w_H)w_H\right)w_K\\
&&+2e^{\alpha\beta}(w_0+w_\Theta)w_K^2\\\nonumber&&+e^\beta(w_0+w_\Theta+w_H)w_K^2;
\end{eqnarray}
The partial partition functions give the contribution to one branch of the total partition function, the total (normalised) partition function conditioned upon the state of the central bond 
is then given by the product of the weight for the left and right branches.
Each of the partial partition functions includes the Boltzmann weight corresponding to the state of the central bond, which is thus counted twice in the full partition function. This double counting is 
corrected by dividing each term by the relevant Boltzmann weight. 
Summing over all the possible states for the central bond gives the total (normalised) partition function, $z$:
\begin{equation}
z=\sum_\sigma \frac{w_\sigma^2}{\lambda_\sigma}.
\end{equation}
In the usual way, the probability of finding a given bond in state $\sigma$ is given by the partition function
conditioned upon this state divided by the total partition function, i.e.
\begin{equation}
p_\sigma=\frac{w^2_\sigma}{z\lambda_\sigma}.
\end{equation}
It should be noted that the density $\rho$ of the walk on the lattice is simply $p_K$.

The free energy per site may be related to $z$ and $q$ through the relation
\begin{equation}
\beta f=\frac{(k-1)\ln z -2\ln q}{2},
\end{equation}
for a full derivation of this expression see~\cite{Pretti:2006qb}.
For the square lattice $k=3$, hence $\beta f=\ln z-\ln q$. 
When multiple solutions to the recurrence relations exist, 
the solution with the lowest free energy is the stable equilibrium solution. 

\section{Results}\label{results}

\begin{figure}
\includegraphics[width=12cm,clip]{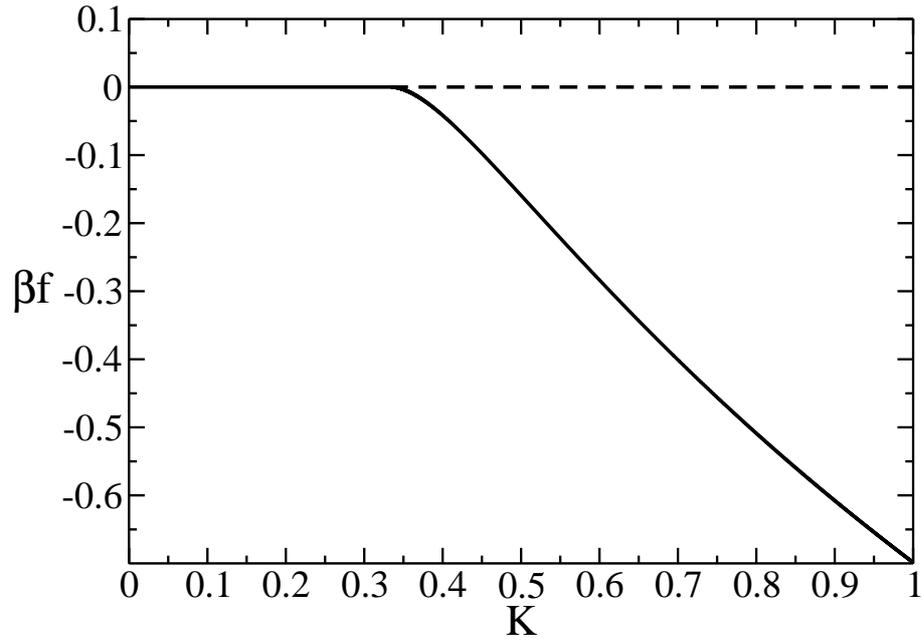}
\caption{$\beta f$ plotted as a function of  $K$ for the $\Theta$-point model ($\alpha=1$) with  $\beta=1.2<\beta_\Theta$. The dashed line shows the meta-stable solution.}\label{theta1}
\end{figure}

It is instructive to see how the calculation works for the pure $\Theta$-point model. Many of the results 
in this case have already been presented by Lise, Maritan and Pelizzola\cite{Lise:1998fj} 
using a different, 
variational, approach to the Bethe approximation. The pure $\Theta$-point model is defined by $\alpha=1$, in which case $\omega_H\equiv 0$ and equation~\ref{rec4} is no longer needed. It is convenient to recast the  relations~\ref{rec1}---\ref{rec3} by setting 
\begin{eqnarray*}
x_K&=&\frac{w_K}{w_0},\\
x_\Theta&=&\frac{w_\Theta}{w_0}.
\end{eqnarray*} 
This leads directly to a trivial solution $x_K=x_\Theta=0$, corresponding to the zero-density phase.
The other possible solutions are given by:
\begin{eqnarray}\label{cubic}
&&x^3_\Theta+(3-e^\beta)x^2_\Theta+(3-2e^\beta)x_\Theta+(e^\beta-1)\left(\frac{1}{3K}-1\right)=0,\\
&&x_K=\sqrt{\frac{3K(1+x_\Theta)^2-1}{3(1+x_\Theta)}}.
\end{eqnarray} 
For a solution to be physically acceptable, $x_K$ and $x_\Theta$ must be positive. Let us first consider the case $1<e^\beta<3/2$. Whilst $K<1/3$ all the coefficients of the cubic equation~(\ref{cubic}) are positive and there is no physically acceptable solution (one of the solutions is negative, and the other 
two are complex conjugates). When $K=1/3$ there is a solution $x_\Theta=0$ and $x_K=0$, and when $K>1/3$ this solution has a negative free energy. This is shown in figure~\ref{theta1}. This corresponds to the high-temperature transition, where the transition point corresponds to an infinite self-avoiding walk, defining $K_\infty(\beta)$ for $\beta<\beta_\Theta$.

\begin{figure}
\includegraphics[width=12cm,clip]{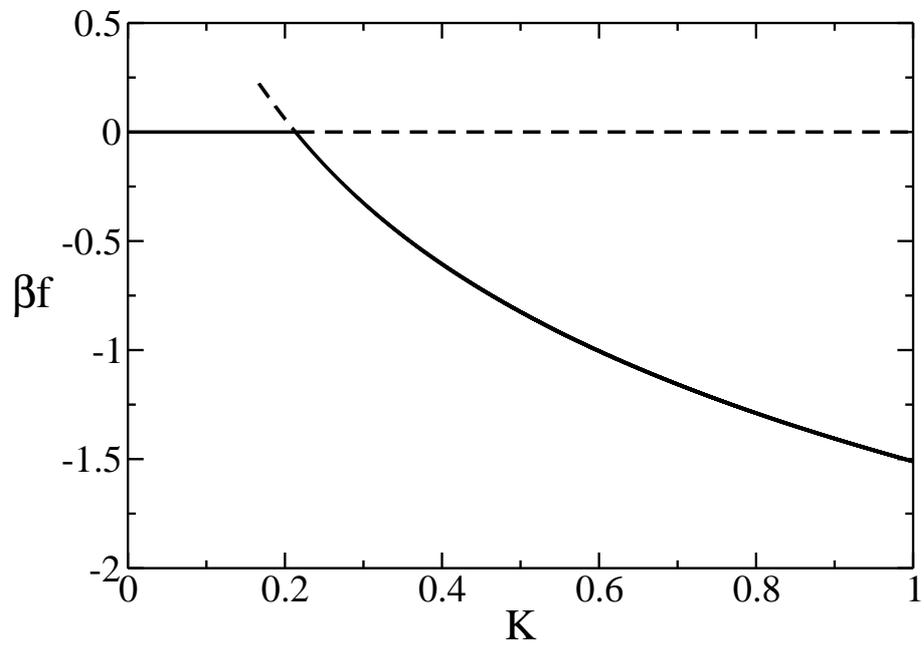}
\caption{$\beta f$ plotted as a function of  $K$ for the $\Theta$-point model ($\alpha=1$) with  $\beta=3>\beta_\Theta$. The dashed line shows the meta-stable solution.}\label{theta2}
\end{figure}

\begin{figure}
\includegraphics[width=12cm,clip]{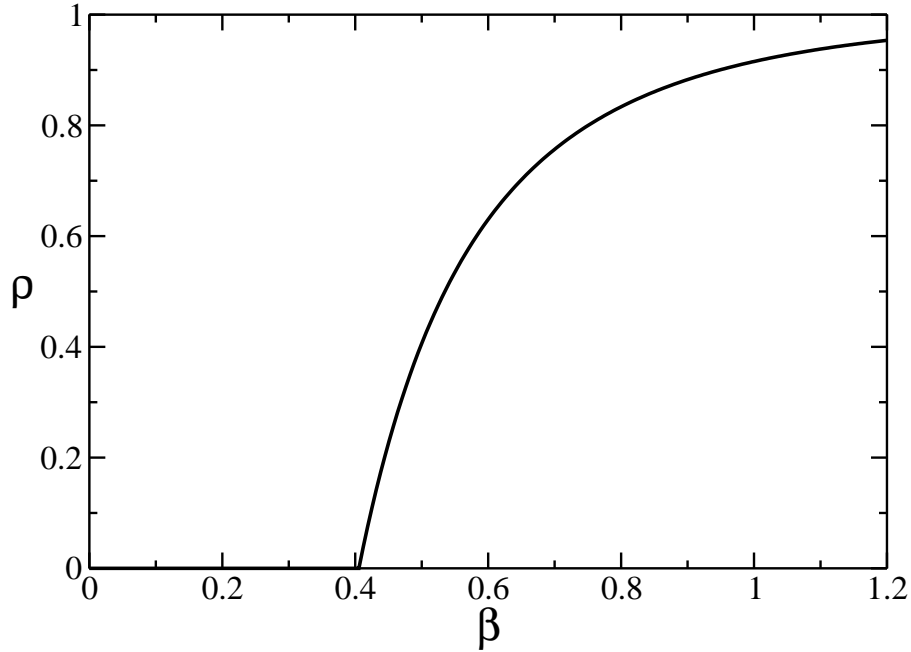}
\caption{Density, $\rho$, plotted along the $K_\infty(\beta)$ line for the $\Theta$-point model ($\alpha=1$).}\label{theta3}
\end{figure}

When $e^\beta=3/2$, the coefficient of the $x_\Theta$ term vanishes,  and when $K=1/3$ the zero-density solution becomes a double root of the cubic equation~(\ref{cubic}). 
This change of behaviour is identified with the $\Theta$-point, and the value $\beta_\Theta=\ln(3/2)=0.405465\cdots$ agrees with the value given by Lise {\it et al}\cite{Lise:1998fj}.
When $e^\beta>3/2$, a physically acceptable solution now exists for values of $K<1/3$. 
To find the transition line for $e^\beta>3/2$ we must check for the stability of this new solution; for small enough $K$ the free energy is positive, and so the solution corresponds to a meta-stable solution, whilst for some higher 
value of $K$, $f$ becomes negative. The point where $f=0$ defines the first-order low-temperature transition, defining $K_\infty(\beta)$ for $\beta>\beta_\Theta$.
The free energy in this case is shown in figure~\ref{theta2} and the density as a function of $\beta$ plotted along the transition line is shown  in figure~\ref{theta3}. 

The tricritical point was identified with a double root of equation~(\ref{cubic}). For $e^\beta\ge 3/2$ there is a line of double roots given by:
\begin{equation}
    K=\frac{9(e^\beta-1)}{4e^{3\beta}}.
\end{equation}
For $\beta>\beta_\Theta$ this gives the position where a second order transition would have occurred had it not been pre-empted by the 
actual first order transition, and as such is identified with the spinodal line.

\begin{figure}
\includegraphics[width=12cm,clip]{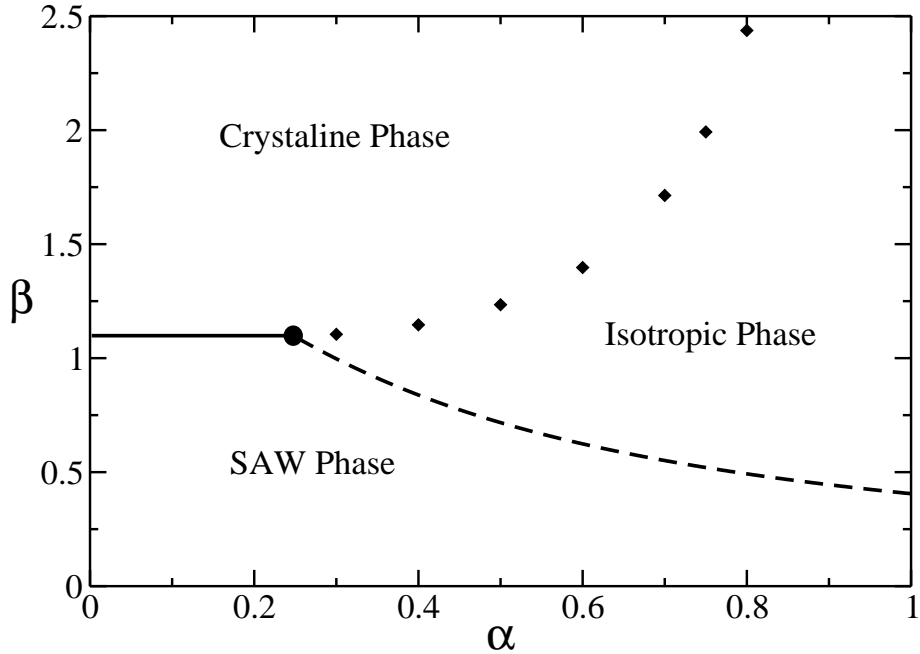}
\caption{The phase diagram in the $\alpha$-$\beta$ plane, with the fugacity, $K=K_\infty(\alpha,\beta)$.
The solid line corresponds to a direct first-order transition between the self-avoiding walk phase and the crystalline phase. The dashed line is the line of $\Theta$-points, separating the self-avoiding walk phase from the isotropic collapsed phase, and the diamonds correspond to the first-order transition between the isotropic and crystalline collapsed phases. The circle shows the location of the multicritical point where the different transition lines meet.}\label{alphapd}
\end{figure}

We now extend our analysis to $\alpha<1$. The analysis follows the same lines as for the pure $\Theta$ model; it is possible to eliminate all the parameters in terms of $x_\Theta$, 
though this is now the solution of a slightly more complicated equation, which may no longer be expressed in a simple polynomial form. The $\Theta$-point extends to a line as $\alpha$ is varied. For  $\beta<\beta_\Theta(\alpha)$ there is still a
zero-density solution for this equation when $K=1/3$, 
corresponding to the self-avoiding walk  transition line. 
The $\Theta$-line is again identified with a double root of equation~(\ref{cubic}) corresponding to the zero-density phase for $K=1/3$, which occurs when 
the coefficient of the $x_\Theta$ term in a small $x_\Theta$ expansion of the equation also vanishes.
This gives the $\beta_\Theta(\alpha)$ through:
\begin{equation}
\alpha=\frac{1}{\beta_\Theta}\ln\left(\frac{27-2e^{\beta_\Theta}}{16}\right).
\end{equation}

\begin{figure}
\includegraphics[width=12cm,clip]{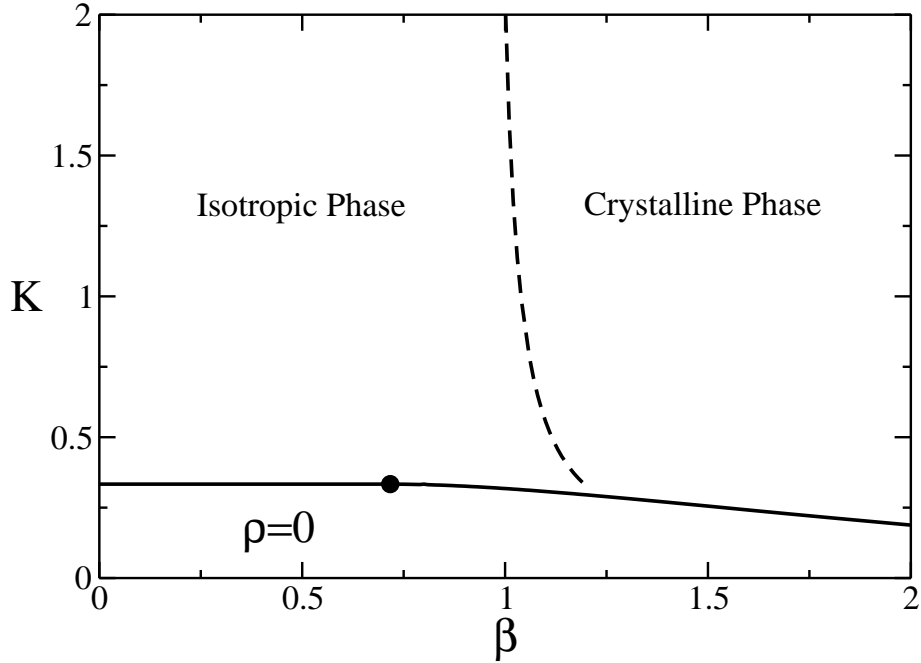}
\caption{The phase diagram in the $\beta$---$K$ plane, with $\alpha=0.5>\alpha_{\rm mc}$.
The circle corresponds to the location of the $\Theta$-point transition. The transition from the $\rho=0$ phase to the crystalline phases is first order, whilst the transition from the $\rho=0$ phase and the Isotropic collapsed phase is second order for $\beta<\beta_\Theta$. The transition between the two dense phases is first order.}\label{pda5}
\end{figure}

\begin{figure}
\includegraphics[width=12cm,clip]{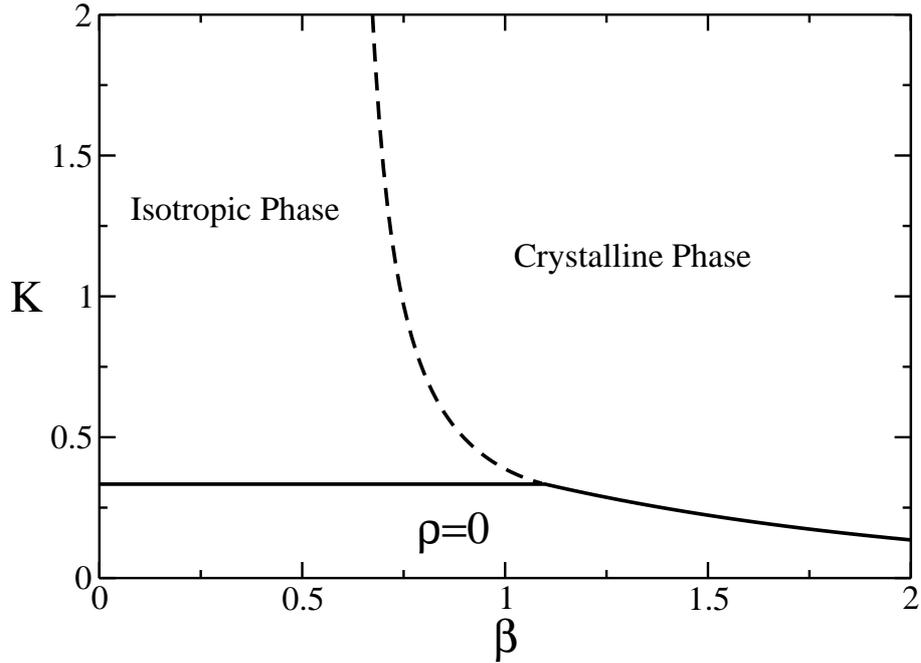}
\caption{The phase diagram in the $\beta$---$K$ plane, with $\alpha=0.2>\alpha_{\rm mc}$.
The transition between $\rho=0$ and the crystalline phase and the transition between the collapsed phases are first order. The $\Theta$ point transition is absent.}\label{pda2}
\end{figure}

For $\alpha<1$ there is the possibility of another phase: the crystalline phase, where the walk fills
the lattice. All the bonds align with one of the lattice directions, maximising the number of
hydrogen bonds. This phase has zero entropy (per lattice site), and its energy per site corresponds to the energy for one bond and one interaction. The free energy for this phase is given by
\begin{equation}
\beta f_{\rm cryst}=-\left(\beta+\ln K\right).
\end{equation}
Setting $K=1/3$, and thus following the self-avoiding walk line, it is seen that $f_{\rm cryst}=0$ when $\beta=\ln 3$. If $\beta_\Theta(\alpha)<\ln 3$, 
the first transition met on increasing $\beta$ 
is the $\Theta$-point transition, and the location of the crystallisation transition, $\beta_H(\alpha)$, 
is determined by comparing the free energies of the collapsed phase and $f_{\rm cryst}$ {\em along the $K_\infty$-line}.
If $\beta_\Theta(\alpha)>\ln 3$, the walk collapses directly to the
crystalline phase, and the $\Theta$-transition is not present. 
The change-over between these two cases occurs at a multi-critical point, which is found by setting
$\beta_\Theta(\alpha_{\rm mc})=\ln 3$. This gives the location of the multi-critical point as: $K_{\rm mc}=1/3, \beta_{\rm mc}=\ln 3\approx 1.0986123$ and $\alpha_{\rm mc}=\ln(21/16)/\ln 3\approx 0.2475247$.
The phase diagram 
projected onto the $K_\infty(\alpha,\beta)$ plane is shown in figure~\ref{alphapd}. When 
$\beta>\beta_H(\alpha)$, the $K_\infty$-line is given by $K=e^{-\beta}$.  
In the high density region of the phase diagram,  comparing the free energies of the collapsed and crystalline phase in the dense region, another phase transition may be seen, already reported 
in the literature\cite{Foster:2008fk,Krawczyk:207qy}. Here this phase transition shows up as a first-order transition, whilst in
more realistic numerical calculations there is evidence to suggest that it is in fact of second order\cite{Foster:2008fk}.
Phase diagrams for two representative cases, $\alpha=0.5$  and $\alpha=0.2$, are shown in figures~\ref{pda5} and~\ref{pda2}.

\section{Discussion}\label{discussion}

Whilst the Bethe approximation is ``only" a mean-field type calculation, 
it usually captures the essential 
features of the model. For $\alpha=1$ we obtain  
results consistent with the results of Lise {\it et al}\cite{Lise:1998fj}.
For $\alpha=0$ the model corresponds to the pure Hydrogen-bonding model. The $(K,\beta)$ phase
diagram, as already remarked by Buzano and Pretti\cite{Buzano:2002hc}, corresponds closely to that found with
transfer matrices\cite{Foster:2001xd}, except that the phase transition between the isotropic dense phase and the crystalline phase is found to be first order here, whilst evidence suggests that it is in fact second order\cite{Foster:2001xd,Foster:2008fk}. In the model studied here we include the effect of the solvent, as compared to the hydrogen type interaction, and this 
enters through the parameter $\alpha$. The solvent-mediated interactions favour a collapse to an isotropic collapsed phase, whilst the hydrogen-bonding interaction tends to align the walk along one of the lattice directions, breaking the rotational symmetry, leading to a crystalline phase. For 
$\alpha$ close to one, the collapse of an infinite chain 
is progressive as $\beta$ is increased. At the collapse transition the fractal 
dimension of the walk is  less than the dimension of the lattice. 
It is expected that the details of the lattice will not influence the transition. The collapse transition is, in this case, in the same universality class as the standard $\Theta$ point. However, once in the dense phase, the dimensions of the walk and the lattice are the same. The walk ``sees" the lattice. 
In the plane where the length of  walk first diverges ($K=K^*(\alpha,\beta)$) we see the appearance of
a second transition, from the isotropic phase to the crystalline phase. 
This transition line extends into a transition plane for $K>K^*(\alpha,\beta)$, as shown in figure~\ref{pda5}.
As $\alpha$ is lowered, a point is reached in which the hydrogen interactions dominate, and the walk collapses directly to a dense crystalline phase. These two regimes are separated by a multicritical point,
$\alpha_{mc}$, where the three transition lines shown in figure~\ref{alphapd} meet. 
The phase transitions found in the context of the Bethe approximation correspond well to the phase
diagrams found numerically for the same model\cite{Foster:2008fk,Krawczyk:207qy}.  Krawczyk et al\cite{Krawczyk:207qy} 
found the transition  between crystalline and isotropic dense phases to be
first order in three dimensions, whilst in two dimensions the order of the transition 
was less clear, and the authors conjectured that the transition is critical. This conjecture
is supported by the numerical study of Foster and Pinettes\cite{Foster:2008fk}. 
Here this transition is found to be first order
due to the mean-field nature Bethe approximation, which should be exact in $d=\infty$ dimensions.

The predicted value of $\alpha_{\rm mc}\approx 0.248$ is close to what is seen numerically ($\alpha_{\rm mc}=0.3\to 0.5$)\cite{Foster:2008fk}. Similar phase diagrams to those found in figure~\ref{pda5} and 
figure~\ref{pda2} are found in other models where frustration effects are important in phase transitions between different dense phases, in particular the vertex-interacting self-avoiding walk\cite{bn89,Foster:2003sh} (figure~\ref{pda2}) 
and the bond-interacting walk\cite{Buzano:2002hc,Foster:2007fk} (figure~\ref{pda5}). 
However, the nature of the transition is found to be very sensitive to details of the interactions. The vertex interacting walk has a phase transition in the dense phase which is in the Ising universality class\cite{bn89,Foster:2003sh} whilst the Hydrogen-bonding model has a transition which is critical, but not Ising ($\nu\approx 0.87$)\cite{Foster:2008fk}. In the context of the Bethe approximation, all these transitions show up as first order. It would be interesting to understand how to incorporate in a 
mean-field type calculation the essential features which would reproduce the second order nature of
the transition between dense phases.

The general features of what is presented here remain true in three dimensions. The self-avoiding walk line occurs for $K=1/5$ rather than $1/3$. This is easily understood: $1/K_{\rm SAW}$ corresponds to the average number of lattice directions available to the walk at each step. Due to the absence of loops on the Bethe lattice, this is simply one less than the co-ordination number of the lattice, i.e. $2d-1$.

\begin{acknowledgments}
The authors would like to thank Claire Pinettes for a careful rereading of the manuscript, and helpful
comments.
\end{acknowledgments}

\end{document}